# Fine-tuning neural excitation/inhibition for tailored ketamine use in treatment-resistant depression

Erik D. Fagerholm[1], Robert Leech[1], Steven Williams[1], Carlos A. Zarate Jr.[2], Rosalyn J. Moran[1,*], Jessica R. Gilbert[2,*]

[1] Department of Neuroimaging, King's College London, London, UK
[2] Experimental Therapeutics and Pathophysiology Branch, NIMH, NIH, MD, USA

Corresponding authors: erik.fagerholm@kcl.ac.uk, jessica.gilbert@nih.gov

* These authors contributed equally to this work

**Abstract**

The glutamatergic modulator ketamine has been shown to rapidly reduce depressive symptoms in patients with treatment-resistant major depressive disorder (TRD). Although its mechanisms of action are not fully understood, changes in cortical excitation/inhibition (E/I) following ketamine administration are well documented in animal models and could represent a potential biomarker of treatment response. Here, we analyse neuromagnetic virtual electrode timeseries collected from the primary somatosensory cortex in 18 unmedicated patients with TRD and in an equal number of age-matched healthy controls during a somatosensory 'airpuff' stimulation task. These two groups were scanned as part of a clinical trial of ketamine efficacy under three conditions: a) baseline; b) 6-9 hours following subanesthetic ketamine infusion; and c) 6-9 hours following placebo-saline infusion. We obtained estimates of E/I interaction strengths by using Dynamic Causal Modelling (DCM) on the timeseries, thereby allowing us to pinpoint, under each scanning condition, where each subject's dynamics lie within the Poincaré diagram – as defined in dynamical systems theory. We demonstrate that the Poincaré diagram offers classification capability for TRD patients, in that the further the patients' coordinates were shifted (by virtue of ketamine) toward the stable (top-left) quadrant of the Poincaré diagram, the more their depressive symptoms improved. The same relationship was not observed by virtue of a placebo effect – thereby verifying the drug-specific nature of the results. We show that the shift in neural dynamics required for symptom improvement necessitates an increase in both excitatory and inhibitory coupling. We present accompanying MATLAB code made available in a public repository, thereby allowing for future studies to assess individually-tailored treatments of TRD.

**Introduction**

Ketamine's rapid antidepressant efficacy was first demonstrated in a clinical trial by (Berman, Cappiello et al. 2000) – a study that lead to intense focus on the glutamatergic system's putative role in mood disorders, including major depressive disorder (MDD) (Zarate, Singh et al. 2006, Murrough, Iosifescu et al. 2013) and bipolar depression (Diazgranados, Ibrahim et al. 2010, Zarate Jr, Brutsche et al. 2012). While still elusive, a mechanistic explanation for ketamine's antidepressant effects could lead to the development of novel, rapid-acting therapeutics or adjunctive treatment options better tailored to individual patients, without unwanted psychoactive side effects and abuse potential. Ketamine, a noncompetitive N-methyl-D-aspartate (NMDA) receptor antagonist (Hirota and Lambert 1996), is thought to exert its anesthetic, dissociative, and psychotomimetic actions via NMDA receptor inhibition (Zanos, Moaddel et al. 2018). However, multiple direct and indirect mechanisms resulting from NMDA receptor inhibition that culminate in increased brain-derived neurotrophic factor (BDNF) release (Lepack, Fuchikami et al. 2014), and $\alpha$-amino-3-hydroxy-5-methyl-4-isoxazolepropionic acid (AMPA) surface expression (Maeng, Zarate Jr et al. 2008, Zanos, Moaddel et al. 2016) have been posited to contribute to its antidepressant actions (Gould, Jr. et al. 2019).

NMDA receptor blockade by subanesthetic dose ketamine of fast-spiking gamma-aminobutyric acid (GABA)-ergic interneurons, particularly the parvalbumin basket cells, leads to local inhibition of interneuron tonic firing and subsequent disinhibition of pyramidal neurons downstream of NMDA receptor antagonism (Homayoun and Moghaddam 2007, Widman and McMahon 2018). This produces an immediate glutamate surge (Moghaddam, Adams et al. 1997), activating the mammalian target of rapamycin complex 1 (mTORC1) pathway (Li, Lee et al. 2010), increasing BDNF release, and activating downstream signaling pathways that stimulate synapse formation (Duman 2014). Recent work has shown that this disinihibition mechanism is shared to varying degrees with other antagonist drugs with rapid-acting antidepressant efficacy, with ketamine being the most reliable (Widman and McMahon 2018) and also demonstrating the most robust and sustained antidepressant effects (Gould, Jr. et al. 2019). In addition, there are a host of cascading intracellular changes following ketamine administration – involving eukaryotic elongation factor 2 – that promote BDNF release (Autry, Adachi et al. 2011, Monteggia, Gideons et al. 2013) and homeostatic synaptic scaling mechanisms (Kavalali and Monteggia 2020). Furthermore, cellular changes result from direct inhibition of extrasynaptic NMDA receptors (Miller, Yang et al. 2014) that activate cellular plasticity mechanisms and promote synaptic potentiation. Given the evidence of direct and indirect changes in cortical excitation/inhibition (E/I) following ketamine administration, deriving noninvasive measures of E/I metrics in treatment-resistant MDD (TRD) participants could serve as useful biomarkers of antidepressant efficacy. Furthermore, E/I metrics have the potential to offer translational opportunities to fine-tune antidepressant response through adjunct pharmacological or neuromodulatory interventions.

We use Dynamic Causal Modelling (DCM) (Friston, Harrison et al. 2003) to describe noninvasively-measured neural dynamics for 18 patients with TRD and an equal number of age-matched healthy control participants, in terms of a coupled E/I model (see Appendix I). These two groups were scanned using magnetoencephalography (MEG) under three conditions: a) baseline, i.e. not under the influence of any pharmaceutical agents; b) post-ketamine, approx. 6-9 hours following intravenous subanesthetic (0.5 mg/kg) ketamine infusion; and c) post-placebo, approx. 6-9 hours following intravenous placebo-saline infusion. Participants were scanned during a somatosensory 'airpuff' stimulation task, and neuromagnetic activity from the primary somatosensory cortex was utilised to examine neural dynamics for each subject and condition. Using the posterior estimates of the DCMs, we were then able to plot the coordinates summarizing each subject's dynamics under each scanning condition within the Poincaré diagram, as defined in dynamical systems theory (see Appendix II).

We show that there is a relationship between the patients' reported depression symptom improvement and the extent to which the ketamine infusion facilitates a shift toward the stable (top-left) quadrant of the Poincaré diagram. This validates the Poincaré diagram as a robust classification tool for TRD ketamine response. Furthermore, we show that this relationship does not arise by virtue of a placebo effect either in TRD patients, or within the healthy control group. Finally, we show that both E and I coupling (rate constants) must increase in order for ketamine to shift patient neural dynamics in the desired direction in the Poincaré diagram (see Appendix III). Future studies may be able to use our techniques to tailor the dosage of ketamine or to add adjunctive pharmacological or neuromodulatory treatments to shift a patient's dynamics in any desired direction in the Poincaré diagram.

**Methods**

**Data collection:** All participants were studied as part of a larger double-blind, crossover clinical trial (NCT#00088699) of ketamine's antidepressant efficacy at the National Institute of Mental Health in Bethesda MD. The present study included 18 participants (10 F/8 M, mean age=36.9±10.7 years) with a DSM-IV-TR diagnosis of MDD without psychotic features, who were experiencing a major depressive episode of at least 4 weeks duration and had a Montgomery Asberg Depression Rating Scale (MADRS) (Montgomery and Åsberg 1977) score of ≥ 20 at screening. They were also treatment-resistant, having failed at least one adequate antidepressant trial as assessed by the Antidepressant Treatment History Form (Sackeim 2001). Healthy control participants (11 F/7 M, mean age=33.9±10.3 years) had no Axis I disorder or family history of Axis I disorder in first-degree relatives. The present study included only those participants who completed all baseline, post-ketamine and post-placebo scan sessions. Additional details on the sample and experimental design have been previously reported (Gilbert, Yarrington et al. 2018).

Neuromagnetic data were collected at 1200 Hz (bandwidth 0-300 Hz) using a CTF 275-channel MEG system with SQUID-based axial gradiometers (VSM MedTech Ltd.). Data were collected during a passive somatosensory stimulation task where participants received tactile stimulation of the right index finger (500 stimuli, 25-millisecond duration, 2-Hz average rate) over a 250-second experimental run. Tactile stimulation was controlled by a pneumatic stimulating device emitting brief bursts of air at 30 psi, displacing a plastic membrane placed against the skin of the distal phalange. During the experimental run, participants were instructed to focus on a stationary fixation dot projected on a screen in front of them. We administered the MADRS at multiple time points to measure change in depression symptomatology for each condition. For baseline scores, we used the ratings collected 60 minutes prior to the first infusion. For ketamine and placebo scores, we used the ratings collected 230 minutes after each infusion. This time point reflects the closest rating collected relative to the MEG recording session. Change in MADRS scores were calculated by subtracting the post-ketamine and post-placebo ratings from the baseline rating – positive scores therefore reflect antidepressant efficacy.

**Data preprocessing:** Offline, MEG data were visually inspected and trials were removed where visible artifacts including head movements, jaw clenches, eye blinks, and muscle movements were present. In addition, individual channels with excessive sensor noise were removed from subsequent analyses. Data were then bandpass filtered from 1-58 Hz and epoched from −100 to 300 ms peristimulus time using analysis routines available in the academic freeware SPM12 (Wellcome Trust Centre for Neuroimaging, http://www.fil.ion.ucl.ac.uk/spm/). Data for each participant and session were coregistered to a canonical template brain and source activity was extracted from left primary somatosensory cortex (-40, -32, 60) with a 5 mm radius using SPM's source-extraction algorithm. Subsequent DCM analysis used this wide-band, 1-58 Hz, 'virtual electrode' signal. For computational efficiency, we used the participant with the least number of trials remaining after data cleaning as the benchmark for subsequent analyses. Thus, we considered the first 389 trials for each participant and session in subsequent analyses. For each recording session, we used the first 121 time points (corresponding to 100 ms) following airpuff stimulation in the DCM analysis. We averaged these segments across all trials for each participant and session in order to obtain mean event-related potential (ERP) timecourses for each patient and control for each of the three scanning sessions.

**Bayesian model inversion:** Following data preprocessing, we performed Bayesian model inversion using the spm_LAP inversion scheme in the SPM software. We use Eq [2] (see Appendix I) as the flow function (equation of motion), together with an extrinsic input to each dependent variable accounting for the influence of the airpuff stimulation and an observer equation consisting of the sum of the E and I dependent variables (see accompanying code). We performed Bayesian model inversion on each of the 108 timeseries (18 subjects × 2 groups × 3 conditions), thereby obtaining posterior estimates for the E and I coupling strengths, from which we calculate a set of coordinates in the Poincaré diagram (see Appendix II). Bayesian model inversion was performed on ERP timecourses after averaging across all available trials for each subject. We show an example of these averaged timecourses for one patient and one healthy control under the three scanning sessions in Figure 1.

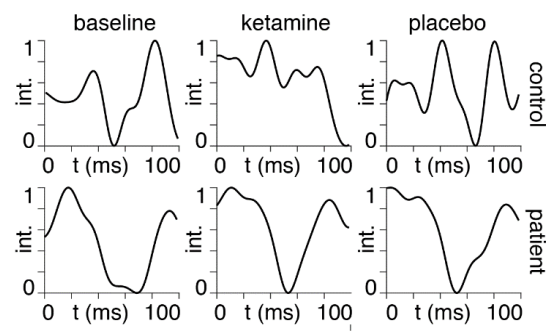

**Figure 1: ERPs.** Averaged ERP timecourses for the first 100ms following airpuff stimulation (first time point). The y-axis shows the MEG signal intensity (int.), normalized between zero and unity. The three conditions: baseline, ketamine and placebo are shown from left-to-right for a healthy control subject (top row) and a single TRD patient (bottom row).

We show mean ERPs for the remaining 34 subjects in Supplementary Figure 1.

**Excitation/inhibition and the Poincaré diagram:** Let us consider a neural region in which the dynamics are described by a balance between E and I (see Appendix I). The Poincaré diagram then allows for an intuitive tool for the visualisation of system dynamics (see Appendix II). We derive the ways in which the E and I coupling strengths must be fine-tuned in order to achieve a shift in the 'southwest' direction in the Poincaré diagram (see Appendix III), as this turns out to be associated with symptom improvement (see Results). Note that, although we focus on this particular shift, it is possible to achieve any arbitrary shift in the Poincaré diagram by choosing the appropriate E/I fine-tuning parameters (see Figure 2).

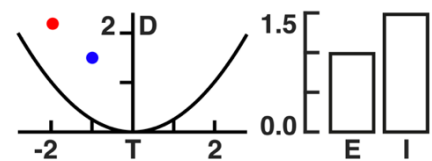

**Figure 2: E/I balance and the Poincaré diagram.** Left: An initial pre-ketamine (blue dot) location and a final post-ketamine (red dot) location in the Poincaré diagram with the Trace (T) on the x-axis and Determinant (D) on the y-axis (see Appendix II). Right: The excitation (E) and inhibition (I) fine-tuning parameters necessary to achieve the shift from the blue to the red dot in the Poincaré diagram on the left (see Appendix III).

We see from Figure 2 that in this case we require that E does not change, as its tuning parameter is equal to unity and we require that I increases by approximately 50% in order to achieve the desired shift.

**Results**

**MADRS scores:** We list MADRS scores for all 18 patients and controls under baseline, ketamine and scanning conditions in Table 1.

**Empirical E/I coupling parameters:** We collect all the posterior E/I coupling parameters recovered by the Bayesian model inversion for patients and controls in the baseline, ketamine, and placebo conditions (see Table 2):

**Table 1: Sub = subject; B = baseline; P = placebo; K = ketamine.** MADRS scores for all patients P01-P18 and controls H01-H18, including the changes in MADRS scores by virtue of ketamine (B-K) and placebo (B-P) infusions. Note that P01 and H01 correspond to the two subjects shown in Figure 1 and all remaining subjects correspond to those shown in Supplementary Figure 1, in the same order.

| Sub | B | P | K | B-K | B-P | Sub | B | P | K | B-K | B-P |
|---|---|---|---|---|---|---|---|---|---|---|---|
| P01 | 31 | 32 | 24 | 7 | -1 | H01 | 0 | 0 | 0 | 0 | 0 |
| P02 | 41 | 35 | 31 | 10 | 6 | H02 | 3 | 2 | 3 | 0 | 1 |
| P03 | 38 | 33 | 20 | 18 | 5 | H03 | 6 | 1 | 0 | 6 | 5 |
| P04 | 25 | 25 | 28 | -3 | 0 | H04 | 1 | 2 | 0 | 1 | -1 |
| P05 | 31 | 29 | 37 | -6 | 2 | H05 | 1 | 1 | 0 | 1 | 0 |
| P06 | 35 | 35 | 11 | 24 | 0 | H06 | 0 | 0 | 1 | -1 | 0 |
| P07 | 26 | 24 | 25 | 1 | 2 | H07 | 0 | 1 | 0 | 0 | -1 |
| P08 | 33 | 24 | 9 | 24 | 9 | H08 | 1 | 1 | 0 | 1 | 0 |
| P09 | 36 | 36 | 41 | -5 | 0 | H09 | 1 | 3 | 0 | 1 | -2 |
| P10 | 30 | 38 | 36 | -6 | -8 | H10 | 0 | 0 | 4 | -4 | 0 |
| P11 | 30 | 30 | 35 | -5 | 0 | H11 | 0 | 0 | 6 | -6 | 0 |
| P12 | 24 | 21 | 10 | 14 | 3 | H12 | 2 | 1 | 2 | 0 | 1 |
| P13 | 38 | 35 | 43 | -5 | 3 | H13 | 4 | 0 | 5 | -1 | 4 |
| P14 | 28 | 29 | 25 | 3 | -1 | H14 | 1 | 0 | 0 | 1 | 1 |
| P15 | 32 | 32 | 25 | 7 | 0 | H15 | 2 | 0 | 1 | 1 | 2 |
| P16 | 35 | 29 | 24 | 11 | 6 | H16 | 4 | 4 | 4 | 0 | 0 |
| P17 | 40 | 41 | 35 | 5 | -1 | H17 | 3 | 1 | 3 | 0 | 2 |
| P18 | 41 | 38 | 36 | 5 | 3 | H18 | 4 | 2 | 3 | 1 | 2 |

**Table 2:** Self-excitatory ($A\_EE$), cross-excitatory ($A\_IE$), cross-inhibitory ($A\_EI$), and self-inhibitory ($A\_II$) coupling parameters (see Appendix I) obtained from the posteriors of the Bayesian model inversion of all 18 controls and patients under baseline, ketamine, and placebo conditions.

| | | BASELINE | | | | KETAMINE | | | | PLACEBO | | | |
|---|---|---|---|---|---|---|---|---|---|---|---|---|---|
| | | A_EE | A_IE | A_EI | A_II | A_EE | A_IE | A_EI | A_II | A_EE | A_IE | A_EI | A_II |
| CONTROLS | | 7.68 | 7.80 | -7.92 | -7.61 | 7.94 | 8.06 | -8.18 | -7.87 | 4.27 | 4.59 | -4.63 | -4.24 |
| | | 0.04 | 0.52 | -0.52 | -0.04 | 5.08 | 5.36 | -5.48 | -5.02 | 1.84 | 2.77 | -2.67 | -1.73 |
| | | 5.44 | 5.68 | -5.72 | -5.41 | 4.73 | 5.02 | -5.06 | -4.70 | 1.02 | 1.38 | -1.48 | -0.98 |
| | | 11.42 | 11.46 | -11.56 | -11.36 | 2.32 | 2.65 | -2.77 | -2.28 | 1.78 | 1.84 | -1.96 | -1.72 |
| | | 8.91 | 8.99 | -9.11 | -8.84 | 0.51 | -0.04 | -0.07 | -0.50 | 9.70 | 9.76 | -9.88 | -9.64 |
| | | 5.34 | 5.57 | -5.66 | -5.29 | 4.21 | 4.48 | -4.60 | -4.14 | 5.48 | 5.72 | -5.84 | -5.42 |
| | | 58.08 | 58.27 | -58.22 | -57.98 | 6.61 | 6.11 | -6.27 | -6.60 | 2.74 | 3.13 | -3.24 | -2.70 |
| | | 7.74 | 7.86 | -7.98 | -7.68 | 7.69 | 7.81 | -7.93 | -7.62 | 5.20 | 5.47 | -5.57 | -5.15 |
| | | 8.26 | 8.36 | -8.48 | -8.20 | 2.90 | 3.37 | -3.44 | -2.87 | 9.17 | 9.25 | -9.36 | -9.10 |
| | | 5.10 | 5.36 | -5.48 | -5.04 | 8.68 | 8.77 | -8.89 | -8.62 | 5.86 | 6.09 | -6.21 | -5.80 |
| | | 3.70 | 4.09 | -4.20 | -3.65 | 2.48 | 3.00 | -3.01 | -2.47 | 0.86 | 0.79 | -0.94 | -0.79 |
| | | 6.64 | 6.83 | -6.87 | -6.61 | 5.44 | 5.68 | -5.72 | -5.40 | 3.39 | 3.78 | -3.89 | -3.34 |
| | | 3.59 | 3.99 | -4.05 | -3.45 | 4.93 | 5.21 | -5.33 | -4.87 | 45.78 | 45.77 | -45.79 | -45.76 |
| | | 6.72 | 6.89 | -7.01 | -6.66 | 20.94 | 20.92 | -20.99 | -20.9 | 0.31 | 0.86 | -0.86 | -0.31 |
| | | 0.55 | 0.77 | -0.77 | -0.54 | 5.40 | 5.64 | -5.69 | -5.36 | -0.13 | 0.31 | -0.31 | 0.13 |
| | | 0.21 | 1.03 | -1.06 | -0.20 | 41.09 | 41.08 | -41.11 | -41.07 | 4.93 | 5.25 | -5.32 | -4.88 |
| | | 6.22 | 6.42 | -6.54 | -6.15 | 4.97 | 5.27 | -5.38 | -4.92 | 8.02 | 8.13 | -8.25 | -7.95 |
| | | 1.06 | 1.61 | -1.61 | -1.05 | 10.06 | 10.12 | -10.23 | -9.99 | 1.70 | 2.21 | -2.22 | -1.69 |
| PATIENTS | | 4.91 | 5.18 | -5.30 | -4.85 | 5.39 | 5.65 | -5.77 | -5.33 | -0.07 | 0.42 | -0.44 | 0.06 |
| | | 1.63 | 1.98 | -2.16 | -1.57 | 2.99 | 3.43 | -3.52 | -2.95 | 8.80 | 8.89 | -9.00 | -8.73 |
| | | 9.48 | 9.54 | -9.66 | -9.41 | 4.62 | 4.83 | -4.95 | -4.56 | 6.41 | 6.60 | -6.72 | -6.35 |
| | | 0.65 | 1.20 | -1.21 | -0.64 | 0.41 | 1.22 | -1.21 | -0.39 | 1.11 | 1.49 | -1.57 | -1.07 |
| | | 2.99 | 3.71 | -3.74 | -2.98 | 6.33 | 6.53 | -6.65 | -6.27 | 15.78 | 15.78 | -15.87 | -15.73 |
| | | 6.77 | 6.95 | -6.99 | -6.73 | 0.62 | 1.11 | -1.11 | -0.61 | 0.39 | 1.32 | -1.34 | -0.39 |
| | | 0.92 | 1.11 | -1.26 | -0.86 | 51.86 | 51.85 | -51.87 | -51.85 | 8.93 | 9.01 | -9.13 | -8.86 |
| | | 5.66 | 5.89 | -6.02 | -5.60 | 0.45 | 0.84 | -0.91 | -0.43 | 56.52 | 56.63 | -56.64 | -56.49 |
| | | 6.82 | 6.98 | -7.11 | -6.76 | 10.74 | 10.78 | -10.89 | -10.68 | 2.77 | 3.11 | -3.21 | -2.73 |
| | | 5.30 | 5.03 | -5.06 | -5.30 | 4.14 | 4.30 | -4.42 | -4.07 | 3.14 | 3.44 | -3.53 | -3.14 |
| | | 8.25 | 8.35 | -8.47 | -8.18 | 4.56 | 4.88 | -4.99 | -4.51 | 1.12 | 1.82 | -1.87 | -1.11 |
| | | 5.83 | 6.04 | -6.12 | -5.78 | 1.95 | 2.53 | -2.55 | -1.94 | 1.98 | 2.64 | -2.55 | -1.79 |
| | | 28.70 | 28.69 | -28.73 | -28.68 | 11.06 | 11.10 | -11.21 | -11.00 | 45.51 | 45.65 | -45.69 | -45.46 |
| | | 1.20 | 0.87 | -0.88 | -1.18 | 0.77 | 1.53 | -1.57 | -0.77 | 30.16 | 30.14 | -30.19 | -30.13 |
| | | 21.04 | 21.09 | -21.09 | -21.03 | 21.17 | 21.16 | -21.22 | -21.13 | 2.72 | 3.07 | -3.19 | -2.66 |
| | | 7.87 | 7.99 | -8.10 | -7.80 | 54.31 | 54.3 | -54.32 | -54.3 | 50.25 | 50.23 | -50.25 | -50.23 |
| | | 4.81 | 5.23 | -5.28 | -4.77 | 5.37 | 5.63 | -5.75 | -5.31 | 9.18 | 9.25 | -9.37 | -9.11 |
| | | 23.20 | 23.18 | -23.24 | -23.17 | 3.01 | 3.25 | -3.37 | -2.96 | 23.93 | 23.63 | -23.66 | -23.83 |

We now make some observations regarding the values in Table 2. To begin with, we note that all four coupling parameters were given priors of zero and not constrained with regard to the sign (±) that they could adopt during the subsequent Bayesian model inversion (see accompanying code). Despite this freedom, with the exception of the fifth control subject in the ketamine condition, all cross-excitatory coupling parameters (A_IE) are positive and all cross-inhibitory coupling parameters (A_EI) are negative, which means that these models are composed of an excitatory and inhibitory component, as per the definitions in Appendix I. The fifth control subject in the ketamine condition has both $A\_IE < 0$ and $A\_EI < 0$, meaning that the best model describing this dataset is one composed of two coupled inhibitory regions, as opposed to a coupled excitatory-inhibitory region as with all others. Disregarding the fifth control subject in the ketamine condition, we note that all but two of the models (control #15 placebo and patient #1 placebo) are self-excitatory $A\_EE > 0$ and self-inhibitory $A\_II < 0$, rendering virtually all models as being composed of a fully-connected E-I system (see Appendix I).

**Plotting in the Poincaré diagram:** We now use the posteriors obtained in Table 2 to plot all models in the space of the Poincaré diagram (see Appendix II) in Figure 3.

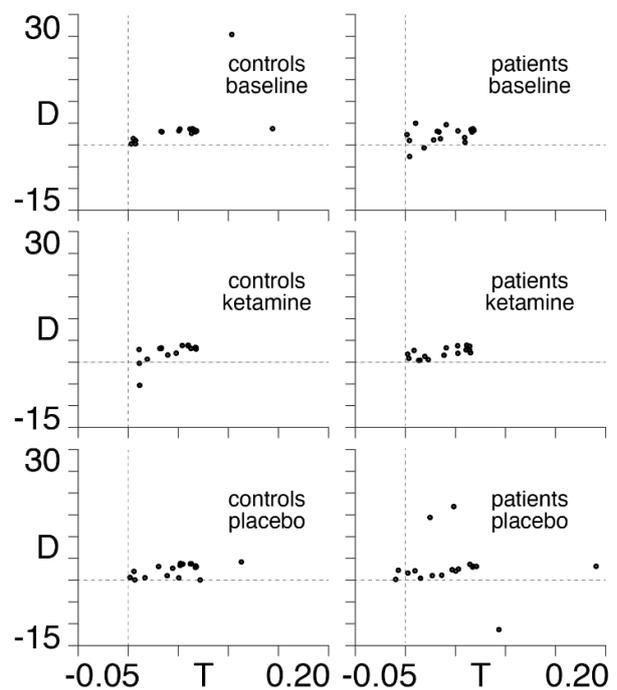

**Figure 3: Results in the Poincaré diagram.** Locations within the Poincaré diagram plotted for all controls (left column) and patients (right column) under baseline (first row), ketamine (middle row) and placebo (last row) conditions. The faint dotted lines show the axes of the trace (T)-determinant (D) plane within the Poincaré diagram.

We note from the top-right subplot in Figure 3 that 16 of the 18 patients in the baseline condition have neural dynamics located in the top-right (unstable) quadrant of the Poincaré diagram.

**MADRS scores and the Poincaré diagram:** In what follows, we perform multiple linear regressions of the changes in MADRS scores (relative to baseline) by virtue of ketamine (B-K in Table 1) and placebo (B-P in Table 1) infusions against associated shifts within the Poincaré diagrams (Figure 4).

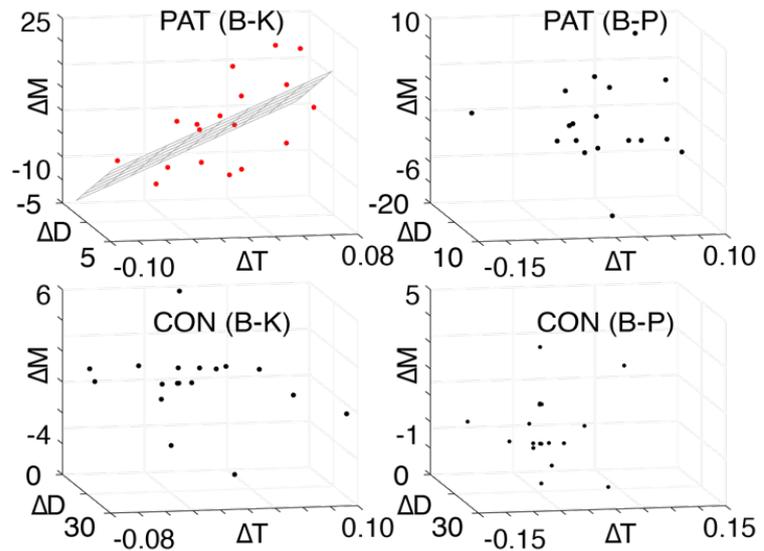

**Figure 4: MADRS scores and the Poincaré diagram.** The changes in the trace ($\Delta T$) and determinant ($\Delta D$) from baseline to ketamine (B-K, first column) and baseline to placebo (B-P, second column) for patients (PAT, first row) and controls (CON, second row), as a function of the associated change in MADRS ($\Delta M$) score. The hyperplane of the multiple linear regression is shown for the only significant model, PAT (B-K), with datapoints in red.

We find that a significant relationship ($p = 0.01, R^2 = 0.43$) is observed for the baseline-to-ketamine conditions in the patient group, with regression coefficients: $\beta_{TR} = 125.1$ and $\beta_{DET} = 1.7$. In other words, we find that symptom improvement is associated with a southwest shift (trace and determinant decrease) in the Poincaré diagram (see Appendix III), with a dominant (99%) westerly component. This, together with the fact that 16 of the 18 patients begin in the northwest quadrant in baseline condition (see Figure 3), demonstrates that symptom improvement is associated with a shift from unstable dynamics (northeast quadrant) to stable dynamics (northwest quadrant) in the Poincaré diagram (see Appendix II). No relationship is observed either in the baseline-placebo condition for the patient group, or for either condition in the control group. Furthermore, upon performing another multiple linear regression of the changes in each of the four individual coupling parameters (see Table 2) between baseline and ketamine conditions in the patient group, we find that this model produces an F statistic of 3.7, as compared with the value of 6.2 obtained for the original trace & determinant model in Figure 4. Therefore, the shifts in trace and determinant provide a better model of the changes in MADRS scores from baseline to ketamine in the patient group than their constituent E/I coupling parameters. This illustrates that the Poincaré diagram representation has explanatory power beyond the individual components of which the trace and determinant are comprised.

**Fine-tuning E/I for optimized treatment response:** Having established the explanatory power of the Poincaré diagram (Figure 4) and its link with E/I balance (see Appendix III), we are now in a position to ask the central question of this paper, namely how should one fine-tune E/I balance in order to optimize treatment response? To answer this, we use the results in Figure 4 – the more the trace and determinant are decreased in patients by virtue of ketamine, the more their symptoms improve. If we perform the associated calculations (see Appendix III), we see that in order for this condition to be satisfied, ketamine must act in a way as to increase both excitation and inhibition relative to the baseline condition.

**Discussion**

Here we demonstrate that network-level excitatory and inhibitory coupling strengths (see Appendix I) can be derived noninvasively from neuromagnetic data, potentially offering an important first step toward personalized, rapid-acting antidepressant treatment for TRD patients. Numerous studies have demonstrated alterations in cortical E/I in MDD (Sanacora, Zarate et al. 2008, Luscher, Shen et al. 2011, Godfrey, Gardner et al. 2018), suggesting that a noninvasive measure of E/I coupling could provide a crucial step in identifying targeted treatments for depression. In addition, antidepressant-dose ketamine has been shown to alter cortical E/I by both directly inhibiting extrasynaptic NMDA receptors (Miller, Yang et al. 2014) and indirectly increasing pyramidal cell excitability, downstream of synaptic GABAergic disinhibition (Homayoun and Moghaddam 2007, Widman and McMahon 2018). These direct and indirect changes in E/I serve to increase BDNF expression (Lepack, Fuchikami et al.

2014), AMPA surface expression (Maeng, Zarate Jr et al. 2008, Zanos, Moaddel et al. 2016), and neuroplasticity-related signaling pathways and synaptic potentiation – mechanisms posited to be foundational to ketamine's antidepressant efficacy (Gould, Jr. et al. 2019).

We have also demonstrated that the Poincaré diagram (see Appendix II) acts as a robust classification tool for predicting efficacy of ketamine in the treatment of TRD. We show that south-westerly (dominantly westerly) shifts in the Poincaré diagram by virtue of ketamine are associated with antidepressant efficacy. This relationship does not exist following placebo saline infusion, suggesting that the Poincaré diagram acts as a biomarker of ketamine antidepressant response. Furthermore, we hypothesize that the mathematical relationships between E/I coupling strengths and associated shifts in the Poincaré diagram (see Appendix III) allow for the potential to individualise adjunctive pharmacological or neuromodulatory antidepressant interventions with ketamine. This relationship could, for example, leverage the downregulation of inhibition through the administration of a GABAergic inhibitor or the upregulation of excitation through repetitive paired-pulse transcranial magnetic stimulation – or even potentially through the titration of drug dosages for the maximization of antidepressant efficacy. Briefly, this involves knowing a target direction in the Poincaré diagram, taken here to be a south-westerly shift from baseline condition (see Figure 4). The beta values in the multiple linear regression indicate a strong (99%) preference for 'west' as the dominant direction. This, together with the fact that the patients in baseline condition are situated in the north-east quadrant of the Poincaré diagram (see Figure 3), means that symptom improvement is associated with the extent of the shift toward the north-west quadrant. This in turn tells us that depressive symptom improvement is associated with the extent to which dynamics are shifted from unstable (north-east quadrant) toward stable (north-west quadrant) dynamics (see Appendix II). Note that the term 'stability' is here used in the context of dynamical systems theory – i.e. a system is said to be stable if it returns to equilibrium following a perturbative influence.

Once a target direction in the Poincaré diagram is known, we can calculate the precise amounts by which E/I coupling parameters need to be fine-tuned in order to optimise the effect of ketamine. We find that optimized shifts within the Poincaré diagram require an increase in both E and I coupling strengths (see Appendix III). It should be noted that these coupling strengths refer to rates of change in both E and I – rather than to E and I themselves. That is to say, an increase in E and I coupling strengths means that we require that the rates at which E and I increase (following a perturbation) must themselves increase. Increased cortical excitability has been previously reported to be associated with antidepressant response to ketamine in a time window overlapping with our measurement window in TRD patients (Cornwell, Salvadore et al. 2012). As drug-induced increases in glutamate levels have been shown to return to baseline within two hours of ketamine administration (Moghaddam, Adams et al. 1997), this lingering increase in cortical excitability is thought to reflect enhanced AMPA receptor glutamatergic neurotransmission (Cornwell, Salvadore et al. 2012, Nugent, Wills et al. 2019). It is interesting to note that increases in inhibitory coupling are also associated with antidepressant response in our sample. Previous studies have demonstrated reduced GABA levels measured by magnetic resonance spectroscopy (MRS) in TRD patients (Hasler, van der Veen et al. 2007, Price, Shungu et al. 2009). In addition, animal work has demonstrated that sustained enhancement of GABAergic transmission is sufficient to elicit antidepressant-like behaviors (Fuchs, Jefferson et al. 2017) and a pilot study of 11 MDD patients found that ketamine administration increased MRS-measured GABA concentrations up to 30 minutes after infusion (Milak, Proper et al. 2016). Taken together, our findings suggest that changes in both excitatory drive (mediated by pyramidal cell disinihibition and subsequent AMPA throughput) and inhibitory drive (mediated by potentiation of GABAergic synaptic inhibition) are important for ketamine's antidepressant efficacy in TRD patients.

While we have demonstrated a methodology to derive fine-tuning parameters for cortical E/I coupling in a TRD patient sample following ketamine administration and a group of healthy control subjects, several limitations should be noted. First, our small TRD patient sample includes participants who have failed at least one antidepressant trial, making this sample relatively homogeneous. Our methodology should be applied to a more heterogeneous sample of depressed patients to determine how robust the derived metrics are for quantifying rapid-acting antidepressant response. Second, the Bayesian Model inversion assumes the most basic linear approximation of a coupled E/I system (see Appendix I), together with an equal balance of E and I contributing to measurement (see accompanying code). Future analyses on larger patient and control groups should be conducted with more complex models including biologically plausible parameters in order to assess trade-offs between complexity and accuracy. Third, several competing theories regarding ketamine's rapid-acting mechanisms of action exist, as previously discussed. While our methods allow one to derive E/I coupling metrics non-invasively, questions remain about the direct relationship between macroscopic changes in E/I coupling and ketamine's cellular and molecular processes and antidepressant mechanisms of action. Finally, the precise extents to which E and I should be increased depends upon the location in the Poincaré diagram towards which neural dynamics should be shifted. In other words, in order to be more specific than 'both E and I should increase', we must identify target coordinates in the Poincaré diagram (see Figure 2). This target could be taken, for example, as the average of a very large cohort of healthy controls in a baseline condition, or a similarly large group of patients that perform particularly well to treatment. Despite these limitations, we have proposed a simple, robust metric for deriving E/I coupling parameters non-invasively on a patient-by-patient basis that shows promise as a potential biomarker of antidepressant efficacy in TRD. Furthermore, we provide a link between neural dynamics, as quantified by coordinates in the Poincaré diagram, and tailored fine-tuning of E/I coupling strengths. We propose that the theoretical foundation, together with the outlined methodology presented here could be used in future clinical trials in order to move toward personalised medicine in the treatment of TRD.

**Appendix I**

**Excitation/inhibition:** Let us suppose that the dynamics in a given neural region can be described in terms of two time-dependent excitatory and inhibitory variables $E$ and $I$ which obey the following two coupled differential equations:

$$\dot{E} = f(E, I)$$
$$\dot{I} = g(E, I), \quad [1]$$

where $f$ and $g$ are non-linear functions.

We can linearize the system in Eq [1] as follows:

$$\dot{E} = A_{EE} E + A_{EI} I$$
$$\dot{I} = A_{IE} E + A_{II} I, \quad [2]$$

where $A_{EE} = \frac{\partial \dot{E}}{\partial E}$, $A_{EI} = \frac{\partial \dot{E}}{\partial I}$, $A_{IE} = \frac{\partial \dot{I}}{\partial E}$, and $A_{II} = \frac{\partial \dot{I}}{\partial I}$ are coupling parameters (see Figure 5).

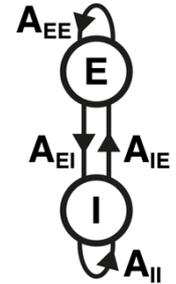

**Figure 5: Excitatory/inhibitory coupling:** An excitatory (E) and inhibitory (I) region, with E-to-E coupling ($A_{EE}$), E-to-I coupling ($A_{EI}$), I-to-E coupling ($A_{IE}$) and I-to-I coupling ($A_{II}$).

The constraints on the signs (whether positive or negative) of the coupling parameters in Eq [2] depend on the specific context of the E/I system under consideration. For instance, in the predator-prey model described by the Lotka-Volterra equations (Lotka 2002), both the self-excitatory ($A_{EE}$) and cross-excitatory ($A_{IE}$) coupling parameters must be positive and the self-inhibitory ($A_{II}$) and cross-inhibitory ($A_{EI}$) coupling parameters must be negative. On the other hand, Turing – in his derivation of the conditions required for spatial inhomogeneities (now known as Turing patterns (Turing 1990)) to emerge in the mixture of a chemical E/I system – stipulated that only the signs of the self-coupling constants be fixed, with $A_{EE} > 0$ and $A_{II} < 0$. Within the context of neural systems, the two coupled differential equations of the E/I Wilson-Cowan model (Wilson and Cowan 1972) each contain a decay term, meaning that both E and

I decrease if left unperturbed. In other words, the self-coupling constants $A_{EE}$ and $A_{II}$ are both negative. For the purpose of all analyses presented here, we stipulate that a pair of variables can be labelled as E and I if their excitatory cross-coupling constant is positive ($A_{IE} > 0$) and their inhibitory cross-coupling constant is negative ($A_{EI} < 0$) – with the self-coupling parameters being allowed to take any sign. That is to say, within a local neural region, the presence of an inhibitory signal will always decrease the amount of excitation and an excitatory signal will always increase the amount of inhibition, but if left uncoupled from one another, E and I are able to both increase and decrease with time.

It is important to keep in mind that when we address 'E/I' balance or fine-tuning strengths, we are refering to the relative sizes of the coupling parameters (rate constants) in Eq [2]. A useful analogy in understanding the E/I balance comes from the classic predator-prey model of a population of chickens and foxes. If put in the same enclosure, the foxes will eat the chickens – i.e. the foxes are an inhibitor with a negative cross-coupling constant – as they decrease the chicken population. On the other hand, the chickens supply food to the foxes – i.e. the chickens are an excitator with a positive cross-coupling constant – as their presence increases the fox population. When we talk about fine-tuning E/I strengths we are referring to changes in these coupling constants – e.g. increasing inhibition means that we increase the rate at which chickens will be eaten by foxes.

**Appendix II**

**The Poincaré diagram:** Eq [2] in Appendix I can equivalently be written in matrix/vector notation as follows:

$$\begin{bmatrix} \dot{E} \\ \dot{I} \end{bmatrix} = \begin{bmatrix} A_{EE} & A_{EI} \\ A_{IE} & A_{II} \end{bmatrix} \begin{bmatrix} E \\ I \end{bmatrix} \quad [3]$$

the Jacobian $J$ of which is given by:

$$J = \begin{bmatrix} A_{EE} & A_{EI} \\ A_{IE} & A_{II} \end{bmatrix} \quad [4]$$

from which we obtain the trace $T$ and determinant $D$ defined as follows:

$$T = A_{EE} + A_{II} \quad [5]$$

$$D = A_{EE}A_{II} - A_{EI}A_{IE}, \quad [6]$$

which we can plot against one another in order to succinctly represent the various dynamical regimes that can be displayed by the system in Eq [2] (Figure 6).

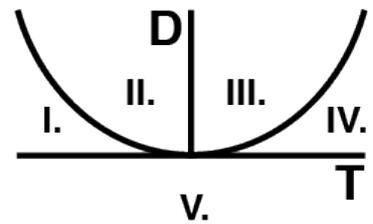

*Figure 6: The Poincaré diagram. The Poincaré diagram, in which we plot the trace $T$ against the determinant $D$, together with a curve defined by the quadratic: $D = \frac{1}{4}T^2$. We therefore divide the plane into the five labeled segments, each of which corresponds to a generic equilibrium: I. Nodal sink; II. Spiral sink; III. Spiral source; IV. Nodal source; and V: Saddle dynamics. Note that only the top-left quadrant (regions I. and II.) are stable – i.e. it is only in this region that dynamics will return to equilibrium following perturbation.*

**Appendix III**

**Excitation/inhibition and the Poincaré diagram:** Let us examine the effect of fine-tuning the elements of the Jacobian in Eq [4] as follows:

$$A_{EE} \to k_E A_{EE}, \quad A_{IE} \to k_E A_{IE} \quad [7]$$

$$A_{II} \to k_I A_{II}, \quad A_{EI} \to k_I A_{EI} \quad [8]$$

where $k_E$ and $k_I$ are positive constants – these are the fine-tuning parameters. Therefore, we assume that ketamine has the effect of fine-tuning the excitatory coupling parameters $A_{EE}$ and $A_{IE}$ by a constant $k_E$ and the inhibitory coupling parameters $A_{II}$ and $A_{EI}$ by a constant $k_I$. The fact that $k_E$ and $k_I$ are positive means that we are able to vary the magnitude (but not flip the signs) of the coupling parameters. In other words, we assume that the fine-tuning of pharmacological agents can result in a shift in E/I balance, but cannot reverse their roles by making E behave like I or vice versa.

We can then derive expressions for the new trace $T_k$ and determinant $D_k$, that result from applying the variations in Eqs [7] and [8] to the original expressions in Eqs [5] and [6]:

$$T_k = k_E A_{EE} + k_I A_{II}, \quad [9]$$

$$D_k = k_E k_I D, \quad [10]$$

Therefore, in order for the dynamics to shift southwest in the Poincaré diagram (see Appendix II) by virtue of the ketamine infusion, we require that $T_k < T$ and $D_k < D$ which, together with Eqs [5], [6], [9], and [10], means that:

$$k_E A_{EE} + k_I A_{II} < A_{EE} + A_{II} \quad [11]$$

$$k_E k_I D < D \quad [12]$$

from which we obtain the following quadratic inequalities for $k_E$ and $k_I$:

$$\frac{A_{EE}}{A_{II}} k_E^2 - \left(\frac{A_{EE}}{A_{II}} + 1\right) k_E + 1 < 0 \quad [13]$$

$$\frac{A_{II}}{A_{EE}} k_I^2 - \left(\frac{A_{II}}{A_{EE}} + 1\right) k_I + 1 < 0 \quad [14]$$

which have the following critical values:

$$k_E = 1 \quad , \quad k_E = \frac{A_{II}}{A_{EE}} \qquad [15]$$

$$k_I = 1 \quad , \quad k_I = \frac{A_{EE}}{A_{II}} \qquad [16]$$

We are therefore now in a position to calculate the ranges of the E/I fine-tuning parameters $k_E$ and $k_I$ required to shift neural dynamics southwest in the Poincaré diagram.

We note from Eqs [15] and [16] that the critical values depend only upon the self-coupling parameters $A_{EE}$ and $A_{II}$. Therefore, as long as $A_{EE}$ and $A_{II}$ have opposite $\pm$ signs, the fact that both $k_E$ and $k_I$ are restricted to positive values means that the inequalities in Eqs [13] and [14] are only satisfied if both $k_E > 1$ and $k_I > 1$. In other words: both E and I must increase.

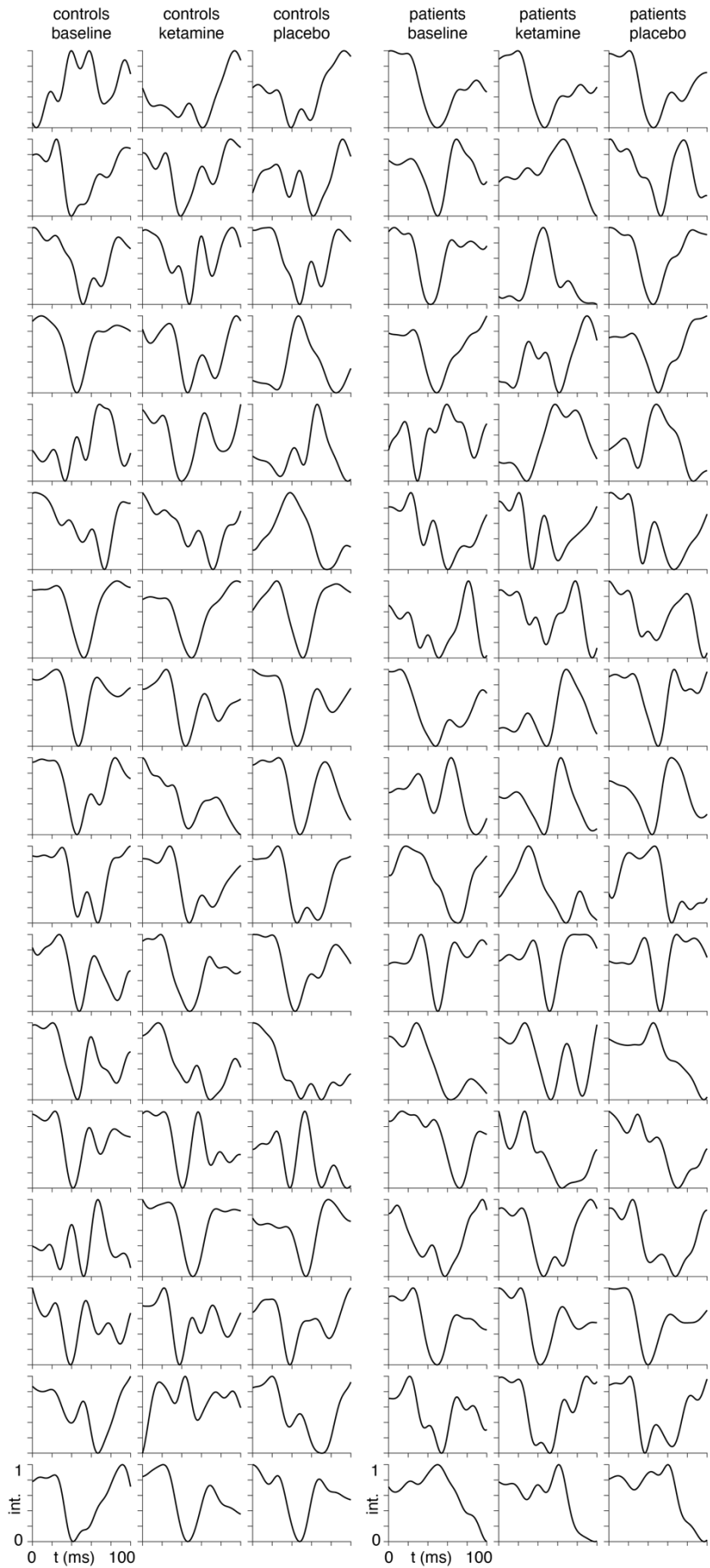

**Supplementary Figure 1: ERPs.** Averaged ERP timecourses for the first 100ms following airpuff stimulation (first time point). The y-axis shows the MEG signal intensity (int.), normalized between zero and unity. The three sessions: baseline, ketamine, and placebo are shown from left-to-right for all the controls (first three columns) and patients (last three columns), excluding the two subjects shown in Figure 1.

**Code availability:** All code is made available in the following public repository: https://github.com/allavailablepubliccode/TRD

**Author contributions:** J.R.G provided the data; all authors designed and performed research, analysed data, and wrote the paper.

**Acknowledgements:** E.D.F. was funded by a King's College London Prize Fellowship; R.L. was funded by the MRC (Ref: MR/R005370/1); R.J.M. and S.W. were funded by the Wellcome/EPSRC Centre for Medical Engineering (Ref: WT 203148/Z/16/Z); C.A.Z., and J.R.G. were funded by by the Intramural Research Program at the National Institute of Mental Health, National Institutes of Health (IRP-NIMH-NIH; ZIA MH002857).

The authors would also like to acknowledge support from the Data to Early Diagnosis and Precision Medicine Industrial Strategy Challenge Fund, UK Research and Innovation (UKRI), the National Institute for Health Research (NIHR), the Biomedical Research Centre at South London, the Maudsley NHS Foundation Trust, and King's College London.

The authors thank the 7SE research unit and staff at the National Institutes of Health for their support.

**Competing interests:** The authors declare no competing interests.